\documentclass[onecolumn,aps,amsmath,amssymb]{revtex4}
\usepackage{mathrsfs}
\usepackage{epsfig,bm,dcolumn}
\usepackage{graphicx}
\usepackage{color}
\usepackage{amsmath}

\begin{document}

\title{Induced interaction in a spin-polarized Fermi gas}

\date{\today}

\author{Zeng-Qiang Yu and Lan Yin}
\email{yinlan@pku.edu.cn}
\address{School of Physics, Peking University, Beijing 100871, China}

\date{\today}

\begin{abstract}

We study the effect of the induced interaction on the superfluid
transition temperature of a spin-polarized Fermi gas.  In the BCS
limit, the polarization is very small in the superfluid state, and
the effect of the induced interaction is almost the same as in the
spin-balanced case.  The temperature $T_t$ and the polarization
$P_t$ of the tricritical point are both reduced from mean-field
results by a factor about 2.22.  This reduction is also significant
beyond the BCS limit.  In the unitary limit, we find
$(P_t,T_t/T_F)=(0.42,0.16)$, in comparison with mean-field and experimental
results.

\end{abstract}



\maketitle

\section{Introduction}
Ultra-cold spin-polarized Fermi gases have attracted a lot of attentions
from both experimental and theoretical sides in the past few years.
In a spin-balanced Fermi gas, BEC-BCS crossover can be achieved at low
temperatures when the scattering between atoms is tuned through
a Feshbach resonance \cite{KetterleReview}, i.e. the
system can evolve smoothly from a BCS superfluid phase to a
molecular-BEC phase \cite{Eagles69,Leggett80}. However, this
crossover picture is no longer accurate with spin polarization, and
phase separation between normal and superfluid phases can take
place at very low temperatures \cite{Hulet06,Ketterle07}.  In addition,
other exotic superfluid phases, such as the FFLO phase
\cite{Fulde64,Larkin64}, may appear \cite{Hulet09}.

The phase separation takes place when the chemical potential difference between two spin species
$h=(\mu_\uparrow-\mu_\downarrow)/2$, reaches a critical value $h_c$.
This first-order transition was first investigated by Clogston and
Chandrasekhar in the context of BCS superconductors
\cite{Clogston62,Chandrasekhar62}.  In the BCS limit, i.e. with a
weakly-attractive interaction, they found that at zero temperature $T=0K$,
$h_c$ is given by $h_c=\Delta/\sqrt{2}$, where $\Delta$ is the energy
gap for $h=0$.  Since the superfluid transition in the spin-balanced
case is a second-order transition, a tricritical point is expected
in the $T-h$ phase diagram.  In the unitarity limit at the resonance, the
tricritical point was observed experimentally \cite{Ketterle08}, with
both the first-order and second-order transition lines located.  In recent
years, the phase diagram of the spin-imbalanced Fermi gas was theoretically studied by mean-field
method \cite{Bedaque03,Yip06,Sheehy06,Hu06,Mueller06},
variational method \cite{Chevy06}, pairing fluctuation theory
\cite{Levin06,Hu06b,Parish07}, Quantum Monte-Carlo simulations
\cite{Pilati08,Lobo06} and renormalization-group approach
\cite{Stoof08}.

In a Fermi gas, the fluctuation in the particle-hole channel
generates the induced interaction which was first pointed out by Gorkov
and Melik-Barkhudarov (GMB) \cite{GMB61}. In a spin-balanced Fermi
gas with a BEC-BCS crossover, it suppresses pairing considerably. In
the BCS limit, the superfluid transition temperature $T_c$ is
reduced from the mean-field value by a factor about 2.22.  The
induced interaction is important in both the BEC limit and unitary
region \cite{Yu09}.  In the BEC limit, the effect of the induced
interaction is negligible due to disappearance of Fermi surface.
The induced interaction has also been studied for a spin-balanced
Fermi gas in optical lattice \cite{Kim09,Heiselberg09} and a homogeneous
three-components Fermi gas \cite{Pethick09}.  In
this paper, we study the induced interaction in a spin-imbalanced
Fermi gas and investigate its impact on the superfluid transition
temperature from the BCS limit to the unitary region.

\section{Induced interaction a spin polarized Fermi gas}

A spin-polarized Fermi gas with a wide Feshbach resonance can be
described by a single-channel model,
\begin{equation}
\mathcal{H}=\sum_{\sigma} \frac{\hbar^2}{2m}|\nabla \psi_{\sigma}|^2
+g\psi_{\uparrow}^{\dagger}
\psi_{\downarrow}^{\dagger}\psi_{\downarrow}
\psi_{\uparrow}-\mu(\psi_{\uparrow}^\dagger \psi_\uparrow
+\psi_{\downarrow}^\dagger \psi_\downarrow)
-h(\psi_{\uparrow}^\dagger \psi_\uparrow-\psi_{\downarrow}^\dagger
\psi_\downarrow), \label{Hamiltonian}
\end{equation}
where the coupling constant is given by $g=4\pi \hbar^2 a_s/m$,
$a_s$ is the scattering length,
$\mu=(\mu_\uparrow+\mu_\downarrow)/2$, $\mu_\sigma$ is the chemical
potential for spin component $\sigma$, and
$h=(\mu_\uparrow-\mu_\downarrow)/2$ is an effective Zeeman field. In
this following, we consider the homogeneous case with spin-up atoms
as the majority component.

In the original work by Gorkov and Melik-Barkhudarov \cite{GMB61},
the induced interaction was obtained in the BCS limit by the
second-order perturbation \cite{Pethick00}.  GMB's treatment can be
extended to the region with a strong interaction beyond the BCS limit
in a random phase approximation (RPA) \cite{Yu09}.  For a scattering
process with $p_1+p_2\rightarrow p_3+p_4$, the induced interaction
is given by
\begin{equation}
U_{\mathrm{ind}}( p_1, p_2; p_3, p_4)= -{g^2\, \chi(p_1-p_4) \over
1+g\chi(p_1-p_4)},
\end{equation}
where $p_{i}=({\bf k}_{i}, \omega_{l_i})$ is a vector in the space
of wave-vector ${\bf k}$ and fermion Matsubara frequency
$\omega_{l}=(2l+1)\pi/(\hbar\beta)$, $\beta=1/(k_BT)$. The
polarization function $\chi$ is given by
\begin{equation}
\chi(p')= {1\over \hbar^2 \beta V}\sum_p
\mathcal{G}_{0\uparrow}(p)\mathcal{G}_{0\downarrow}(p+p')
=\int {\rm{d}^3{\bf k}\over (2\pi)^3} {f_{{\bf
k}\uparrow}-f_{{\bf k}+{\bf k'}\downarrow}\over i\hbar
\Omega_l+\xi_{{\bf k}\uparrow}-\xi_{{\bf k}+{\bf k'}\downarrow}},
\nonumber
\end{equation}
where $p'=({\bf k}', \Omega_{l})$, $\Omega_{l}=2l\pi/(\hbar\beta)$
is the Matsubara frequency of a boson, $V$ is the volume, $f_{\bf k
\sigma}=1/[1+\exp(\beta \xi_{\bf k \sigma})]$ is the Fermi
distribution function, $\xi_{\bf k \sigma }=\epsilon_{\bf
k}-\mu_\sigma$, and $\epsilon_{\bf k}=\hbar^2k^2/2m$.  The Matsubara
Green's function of a non-interacting Fermi gas is given by
$\mathcal{G}_{0\sigma}(p)=\hbar/(i\hbar\omega_l-\xi_{\bf k\sigma})$.

Including the induced interaction, the effective pairing interaction
between atoms with different spins is given by
\begin{equation}
U_{\mathrm{tot}}( p_1, p_2; p_3, p_4)=g +U_{\mathrm{ind}}( p_1,
p_2; p_3, p_4)={g \over 1+g\chi(p_1-p_4)}.
\label{induce}
\end{equation}
Although the effective interaction is a function of both momentum
and frequency, only its s-wave part plays an important role on
pairing at low temperatures. As in GMB's work, we approximate this
s-wave component $g'$ by averaging the polarization function
$\chi_s=\langle\chi\rangle$,
\begin{equation}
g'={g \over 1+g\chi_s}. \label{s-wave}
\end{equation}
In this work, the possibility of FFLO state is ignored and we only
consider pairing between atoms with opposite momentum.  The average of the
polarization function $\chi_s$ is obtained by setting
the frequencies to zero, setting wavevectors to ${\bf k}_1=-{\bf k}_2$,
${\bf k}_3=-{\bf k}_4$, $k_1=k_2=k_3=k_4=k_F$,
and integrating over the angle $\theta$ between ${\bf k}_1$ and ${\bf k}_4$,
\begin{equation}
\chi_s={m\over 8\pi^2\hbar^2}\int_{-1}^{1}{\rm
d}\cos\theta   \int_0^\infty{\rm d} k\, {k\over
k'}[f_{k\uparrow}\ln \left|{k'^2-2kk'+4mh\over k'^2+2kk'+4mh}\right| +f_{k\downarrow}\ln \left|{k'^2-2kk'-4mh\over
k'^2+2kk'-4mh}\right|] ,\label{average_chi}
\end{equation}
where $k_F=(3\pi^2n)^{1/3}$, $n$ is the
total density, and the variable $k'$ is a function of
$\theta$, $k'=|{\bf k}_1-{\bf k}_4|=k_F\sqrt{2(1+\cos\theta)}$. When
$h=0$, Eq. ({\ref{average_chi}) recovers the result in spin-balanced
case \cite{Yu09}.  As indicated by Eq. (\ref{s-wave}),
the effective s-wave interaction is determined by
the average of the polarization function $\chi_s$.

In the BCS limit, $k_Fa_s\rightarrow 0^-$, the superfluid transition
takes place at temperatures much lower than the Fermi temperature
$T_F$.  In this zero-temperature limit, the average of the polarization
function $\chi_s$ is given by
\begin{widetext}
\begin{align}
\chi_s=-{m\over 8\pi^2\hbar^2}{1\over
k_F^2}\int_{0}^{2k_F}{\rm d}k' & \Big{[}\left({k_{F\uparrow}^2\over
2}-{(k'^2+4mh)^2\over 8k'^2}\right)\ln
\left|{k'^2+2k_{F\uparrow}k'+4mh\over
k'^2-2k_{F\uparrow}k'+4mh}\right| +{k'^2+4mh\over 2k'}k_{F\uparrow}
\nonumber \\+ &\left({k_{F\downarrow}^2\over 2}-{(k'^2-4mh)^2\over
8k'^2}\right)\ln \left|{k'^2+2k_{F\downarrow}k'-4mh\over
k'^2-2k_{F\downarrow}k'-4mh}\right| + {k'^2-4mh\over
2k'}k_{F\downarrow}\Big{]}, \nonumber
\end{align}
\end{widetext}
where $k_{F\uparrow}=\sqrt{2m\mu_\uparrow}/\hbar$ and
$k_{F\downarrow}=\sqrt{2m\mu_\downarrow}/\hbar$.
For given total density $n=n_\uparrow+n_\downarrow$, the average
$\chi_s$ is shown as a function of spin polarization
$P=(n_\uparrow-n_\downarrow)/(n_\uparrow+n_\downarrow)$ in Fig.
\ref{fig_chi}. When $P=0$, GMB's result for the spin-balanced case,
$$\chi_{s0}=-{1\over3}\ln(4e)N(E_F),$$ is recovered, where
$N(E_F)=mk_F/(2\pi^2\hbar^2)$ is the density of states at Fermi
energy. As the polarization $P$ increases, the absolute value of
$\chi_s$ decreases, indicating that the induced
interaction becomes weaker.  The average $\chi_s$
varies very slowly for small polarization, for example
$\chi_s=0.96\chi_{s0}$ at $P=0.5$.  For a
nearly full-polarized system, $P\rightarrow 1$, the particle-hole
fluctuation still exists with a non-zero $\chi_s$. In
this limit, the system is described by the picture of Fermi polarons
\cite{Lobo06}, rather than the superfluid transition.

\begin{figure}
\includegraphics[width=8cm]{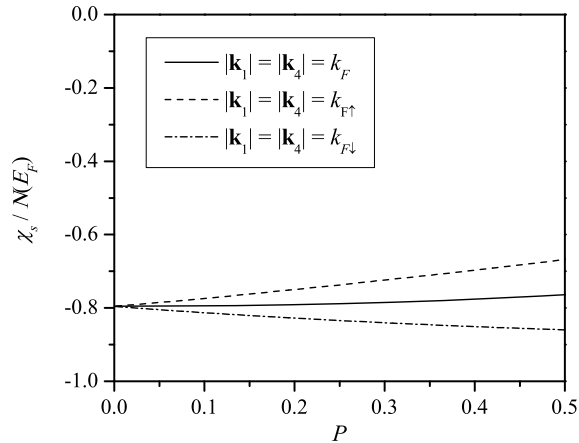}
\caption{ Average of the polarization function $\chi_s$
versus spin polarization $P$ at zero temperature.  The solid line
is obtained by setting wavevectors at $k_F$, the dashed line is
computed at $k_{F\uparrow}$, and the dotted-dashed line is computed at $k_{F\downarrow}$.
When $P \ll 1$, there is very little difference among these three results.
The effective s-wave interaction is given by
${g'}^{-1}=g^{-1}+\chi_s$.}
\label{fig_chi}
\end{figure}

It is worth to note that besides the method described above, we can
use other methods to compute the average of the polarization
function $\chi_s$.  Since the dominant contribution
to pairing comes from atoms near the two Fermi surfaces,
$k_{F\uparrow}$ and $k_{F\downarrow}$, the average
$\chi_s$ can also be computed at wave-vectors
$|{\bf k}_1|=|{\bf k}_4|=k_{F\uparrow}$ or
$|{\bf k}_1|=|{\bf k}_4|=k_{F\downarrow}$,
which are not significantly different for $P\lesssim 0.45$, as shown in Fig. \ref{fig_chi}.
In the spin-balanced limit, $P\rightarrow 0$, these methods produce
the same result.  Even for $P \approx 0.45$, the quantitative difference between
$\chi_s$ taken $k_{F\uparrow}$ or $k_{F\downarrow}$ and
$\chi_s$ taken $k_F$ is less than $10\%$.  As shown in the next section,
the polarization $P$ is less than $0.42$ at tricritical points on the BCS side.
In the following, the average of the polarization
function $\chi_s$ is taken at $k_F$.

\section{Superfluid transition}
We replace the coupling constant $g$ by the effective s-wave interaction $g'$
from Eq. $(\ref{s-wave})$.
When pairing fluctuations are ignored, the thermodynamic potential
is given by
\begin{equation}
\Omega= -{1\over \beta V}\sum_{\bf k}\left[ \ln(1+e^{-\beta
(E_k-h)}) + \ln(1+e^{-\beta (E_k+h)}) \right]
+{1\over V} \sum_{\bf k}(\xi_k-E_k)  -{|\Delta|^2 \over g'}
,\label{Omega}
\end{equation}
where $E_k=\sqrt{\xi_k^2+|\Delta|^2}$ is the energy of the
quasi-particle,
$\Delta=g'\langle\psi_\downarrow\psi_\uparrow\rangle$ is the order
parameter, $\xi_k=\epsilon_k-\mu$. The difference between Eq.
(\ref{Omega}) and the mean-field expression of the thermodynamic potential
is that the particle-hole
fluctuation has been taken into account through the effective interaction $g'$.

The second-order superfluid transition is determined by the condition
\begin{equation}
\lim_{\Delta\rightarrow 0}{1\over \Delta}{\partial \Omega \over
\partial \Delta^* }=0,  \label{st-condition}
\end{equation}
which yields
\begin{equation}
{m\over 4\pi\hbar^2 a_s}+\int{{\rm d}^3k\over (2\pi)^3}
\left[{1-f_{k\uparrow}-f_{k \downarrow} \over 2\xi_k}-{1\over
2\epsilon_k}\right] +\chi_s =0, \label{Tc-eq}
\end{equation}
where the last term in the integrand on l.-h.-s. of this equation is a
counter term due to vacuum renormalization.
If $\mu_\uparrow=\mu_\downarrow$, Eq. (\ref{Tc-eq}) is just the
$T_c$-equation for the unpolarized Fermi gas, which in the BCS limit
produces the superfluid transition temperature in GMB theory  \cite{GMB61},
\begin{equation}\label{Tc}
T_c={\gamma \over \pi}\left({2 \over e}\right)^{7/3} T_F \exp\left({\pi \over 2k_Fa_s}\right).
\end{equation}

In a spin-polarized Fermi gas, the superfluid transition is a first-order
phase transition at very low temperatures.   The first-order and second-order
phase transition lines meet at a tricritical point $(h,T)=(h_t,T_t)$.
At the tricritical point, in addition to Eq. (\ref{st-condition}), we have
\begin{equation}
\lim_{\Delta\rightarrow 0}{1\over \Delta^2}{\partial^2 \Omega \over
\partial {\Delta^*}^2} =0. \label{tricritical-condition}
\end{equation}
In Landau's theory about phase transition, these two equations determine
zero points of the first two coefficients in expansion of
thermodynamic potential in terms of the order parameter.  Eq.
(\ref{tricritical-condition}) can be explicitly written as
\begin{equation}
\int{{\rm d}^3 k\over (2\pi)^3}\left[{1-f_{k\uparrow}-f_{k
\downarrow} \over \xi_k^3} +\beta { {\rm sech}^2 {\beta
\xi_{k\uparrow}\over 2}+ {\rm sech}^2 {\beta \xi_{k\downarrow}\over
2}\over 4\xi_k^2}\right]=0. \label{inflection}
\end{equation}
For a given averaged-chemical-potential $\mu$, the tricritical point $(h_t,
T_t)$ can be determined from coupled equations (\ref{Tc-eq}) and
(\ref{inflection}).

In the BCS limit, both $T_t $ and $h_t$ are proportional
to the superfluid transition temperature $T_c$ given by Eq. (\ref{Tc}) for
the spin-balanced Fermi gas with the same total density and scattering length.
From Eq. (\ref{inflection}), we obtain
\begin{equation}
h_t=1.911k_B T_t, \label{ht}
\end{equation} which leads to
\begin{equation}
P_t= {3h_t \over 2 E_F}=2.867{T_t\over T_F}.
\end{equation}
Putting Eq. (\ref{ht}) into Eq. (\ref{Tc-eq}), we have
\begin{equation}
T_t =0.561T_c.
\end{equation}
Since the induced interaction reduces $T_c$ from the mean-field result by a factor
about $2.22$, both temperature and polarization at the tricritical point are also
reduced by the same factor,
\begin{align}
  {T_t} &=0.561T_c=0.156 T_F e^{\pi/2k_Fa_s}, \label{Tt_limit} \\
  {P_t} &=1.608T_c/T_F=0.446 e^{\pi/2k_Fa_s}. \label{Pt_limit}
\end{align}
This result is the same if we choose to compute the average of the polarization
function $\chi_s$ at $k_{F\uparrow}$ or $k_{F\downarrow}$, because
the polarization $P_t$ approaches zero in the BCS limit.

\begin{figure}
\includegraphics[width=8cm]{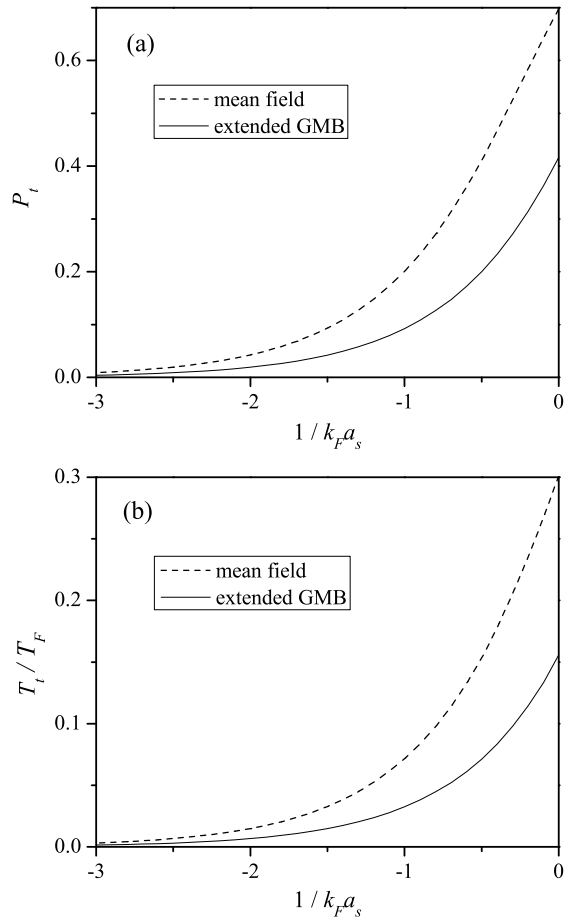}
\caption{ Tricritical point in a polarized Fermi gas as a
function of $1/k_Fa_s$: (a) tricritical polarization $P_t$, and (b)
tricritical temperature $T_t$.  The solid lines are our results
obtained in the extended-GMB approach, and the dashed lines are predictions from
the mean-filed theory.} \label{fig_tricritical}
\end{figure}

Numeric solutions of tricritical polarization $P_t$ and
tricritical temperature $T_t$ beyond the BCS limit are showed in
Fig. \ref{fig_tricritical} as a function of $1/k_Fa_s$.   Both $P_t$ and $T_t$ are reduced
significantly from the mean-field results due to the induced interaction.
At unitary, we obtain  $(P_t,T_t/T_F)=(0.42,0.16)$, considerably smaller than
mean-field results $(P_t,T_t/T_F)=(0.70,0.30)$, but still larger than experimental
result $(P_t,T_t/T_F)\simeq(0.20,0.08)$ \cite{Ketterle08}.  The
discrepancy between experimental and our results are probably due to pairing fluctuations
ignored in our approach which are important in the unitary region and on the BEC side.
In the renormalization-group approach \cite{Stoof08} with both particle-hole and particle-particle
scattering considered, the tricritical point was obtained close to the experimental result.

\section{Discussion and Conclusion}

In a spin-balanced fermi gas, pairing fluctuations can be considered in
the approach pioneered by Nozi\`{e}res and Schmitt-Rink (NSR) \cite{NSR85}. However, in a
spin imbalanced Fermi gas, simple applications of NSR theory failed in the unitary region
\cite{Hu06b,Parish07}.  Hence, a more sophisticated consideration of
this problem is needed in the future.  Another interesting issue is
how to generalize the induced interaction to the broken symmetry
state where the average polarization function $\chi_s$ can
be quite different from the present form due to the large pairing gap $\Delta$ and
the effect of the induced interaction may be more complicated.

In conclusion, we study the effect of the induced interaction on the superfluid
transition temperature in a spin-polarized Fermi gas with a wide Feshbach resonance.
In the BCS limit, the absolute value of the induced interaction
decreases as the polarization $P$ increases, but this change is very small for $P<0.5$.
Both temperature and polarization at the tricritical point are reduced from mean-field
results by a factor $2.22$.  Beyond the BCS limit, reductions of the tricritical polarization $P_t$ and
tricritical temperature $T_t$ are also significant.  In the unitary limit, the tricritical point
is found at $(P_t,T_t/T_F)=(0.42,0.16)$, and the discrepancy with the experimental result indicates
the importance of pairing fluctuations at unitarity.

\section*{ACKNOWLEDGMENTS}

This work is supported by NSFC under Grant No.
10674007 and 10974004, and by Chinese MOST under grant number 2006CB921402.

\end{document}